\documentstyle[11pt,newpasp,twoside,epsf]{article}
\newcommand{\mdot}{{\rm \dot{M}}}

\newcommand{\msun}{{\rm M_{\odot}}}

\newcommand{\lambdabar}{                          
  \hbox{\raise 0.1em
  \hbox to 0pt {--\hss}$\lambda$}}

\renewcommand{\epsilon}{\varepsilon}              

\newcommand{\pspin}{P_{\rm spin}}
\newcommand{\porb}{P_{\rm orb}}

\newcommand{\rmag}{R_{\rm mag}}

\def\edcomment#1{\iffalse\marginpar{\raggedright\sl#1\/}\else\relax\fi}
\reversemarginpar

\begin{document}
\title{Thermal-timescale mass transfer and magnetic CVs}
 \author{Klaus Schenker, Graham A.~Wynn \& Roland Speith}
\affil{Theoretical Astrophysics Group, Department of Physics and
Astronomy, University of Leicester, University Road, Leicester
LE1~7RH, U.K.}

\begin{abstract}
    We investigate the spin evolution of the unusual magnetic CV
    AE~Aqr. As a prototype for a potentially large population of 
    CVs subject to a thermally unstable phase of mass
    transfer, understanding its future is crucial. We present a 
    new definition of the magnetospheric
    radius in terms of the white dwarf's spin period, and use this
    along with numerical simulations to follow the spin evolution 
    of AE~Aqr. We also present preliminary SPH results suggesting
    the existence of a stable propeller state.
    These results highlight the complexity 
    of mCVs and may provide am improved understanding of the
    evolution of all types of CVs.

\end{abstract}


\section{Introduction}

The evolutionary relations among subclasses of magnetic and
non-magnetic cataclysmic variables (mCVs and CVs respectively) 
have been studied for quite
some time. Nevertheless, our understanding of these relations
remains incomplete, caused largely by serious 
selection effects which prevent completeness of the samples.
Here we consider the future evolution of the peculiar mCV
AE~Aqr. Its future spin evolution and equilibrium
states could clear much of the current confusion, as this
binary has been identified as member of a potentially large group of 
post-thermal timescale mass transfer systems (Schenker et al., 2002). 
One consequence of this is the prediction of a relatively large 
population of mCVs at long orbital periods. Such a population 
has not been detected. We consider the possibility that this 
could be because these long period mCVs occupy spin equilibria which
significantly hamper accretion onto the white dwarf (WD),
causing them to have been overlooked or misclassified.

\begin{figure}
    \plotone{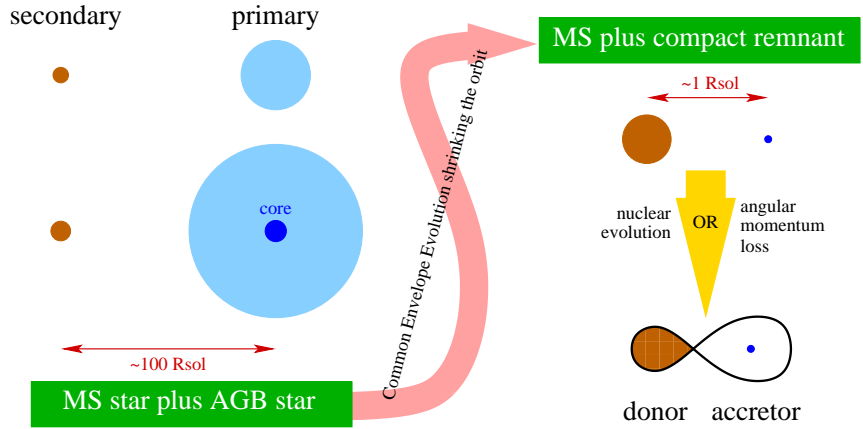}
    \caption{Schematic for the formation of CVs. Note that even
      systems brought into contact by nuclear evolution in the pre-CV
      phase (right column) can later turn into angular momentum loss
      driven semi-detached binaries with a red dwarf donor. If
      (in contrast to the sketch) the donor is initially more massive
      than the WD, thermal-timescale mass transfer ensues.} 
\end{figure}

\subsection{Thermal-timescale mass transfer}

We first consider the 
origin of CVs (Fig.~1). The progenitor of the WD starts off
as the more massive star in a relatively wide 
binary, which has evolved through various stages of (single) stellar 
evolution forming a degenerate core in its centre. During the AGB
phase the star swells and fills its Roche lobe. A subsequent
phase of common envelope evolution leaves the system with the exposed
core (pre-WD) and the largely unaffected secondary in a close
orbit. In order to be a future CV this system has to become
semi-detached. This can be achieved by angular momentum loss 
(AML) shrinking the orbit or nuclear evolution increasing the 
secondary's radius.
The latter channel can
lead to a large fraction of systems with a secondary more massive than
the WD primary leading to thermal-timescale mass transfer (TTMT). 


Examples of TTMT evolution
are the various model tracks for AE~Aqr presented in Schenker et al.\
(2002).
In contrast to standard CVs (which have low mass
donors and stable, AML-driven mass transfer) this
subclass passes through an initial phase of high mass transfer 
prior to become an AML-driven CV.
If we consider CVs which have suffered TTMT, we expect there to be 
differences to standard CVs at the onset of mass transfer:
(i) The WDs are likely to have large rotation rates and masses. 
This is a consequence of the high mass transfer rates
during the supersoft phase (during TTMT) which allows 
significant accretion onto the WD.
(ii) Instead of low-mass MS stars, the donors are
the evolved cores of more massive stars. Their different
internal structure, manifesting itself e.g.\ in a different
mass-radius exponent $\zeta$, leads to smaller $M_2$ at the
same $P_{\rm orb}$, and even to different mass transfer rates.
(iii) The magnetic fields on either or both the donor and the
WD may be different (cf.\ Cumming, 2002, and this volume, on
${\rm B}$--field suppression in rapidly accreting WDs).

\subsection{Non-synchronous mCVs}

The majority of known mCVs (Fig.~2) are either polars
(mostly below the period gap) or intermediate polars (IPs, above the
gap up to $\sim 6$ h). There are only a few, mostly peculiar systems
longwards of $6$ h, one of which is AE~Aqr. If many other systems
like it have passed through a similar phase, where do they appear on
the period distribution? 
This lack of magnetic systems between 6 and 10 h we termed the 
``IP gap'', well aware of the sparse observational data to establish
its existence (cf.\ recently reheated discussions on evidence for a
period gap in mCVs).

\section{White Dwarf spin and magnetosphere}

\begin{figure}
    \plotone{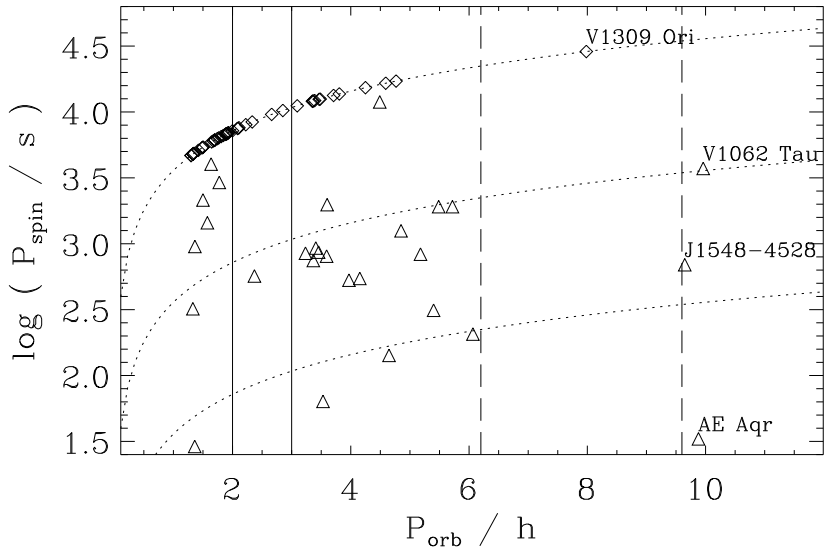}
    \caption{White dwarf spin vs.\ orbital periods of magnetic CVs. Diamonds and
      triangles mark polars and intermediate polars respectively (from
      Ritter \& Kolb, 1998, with various updates). The dotted curves
      indicate $\pspin / \porb = 1$, $0.1$, and $0.01$ from top to
      bottom. The CV period gap between 2 and 3 h is shown by a pair
      of full vertical lines, whereas dashed lines surround a region
      almost devoid of mCVs (``IP-Gap'').}
\end{figure}

We use a simplified description for the magnetic drag on gas streaming
past a spinning WD (King, 1993), where the magnetic acceleration
$a_{\rm mag}$ is proportional to the velocity shear
$\left( v_{\rm B-field} - v_{\rm flow} \right)$.
After splitting off the radial power dependency in the drag and
replacing the ${\rm B}$--field and flow velocities with the WD
rotation and Keplerian motion respectively we obtain
\begin{equation}
    a_{\rm mag} = k_0 \, R^n \, 
    	( R \, \Omega_{\rm spin} - \sqrt{{{\rm G} \, M_1}/{R}} )
    \, .
\end{equation}
Different interaction models are reduced to variations in the 
drag coefficient $k_0$ and the radial power $n \leq 0$.
Propelling and accreting states are possible depending on
the WD spin, which determines the sign of the velocity shear.

\begin{figure}
    \plotone{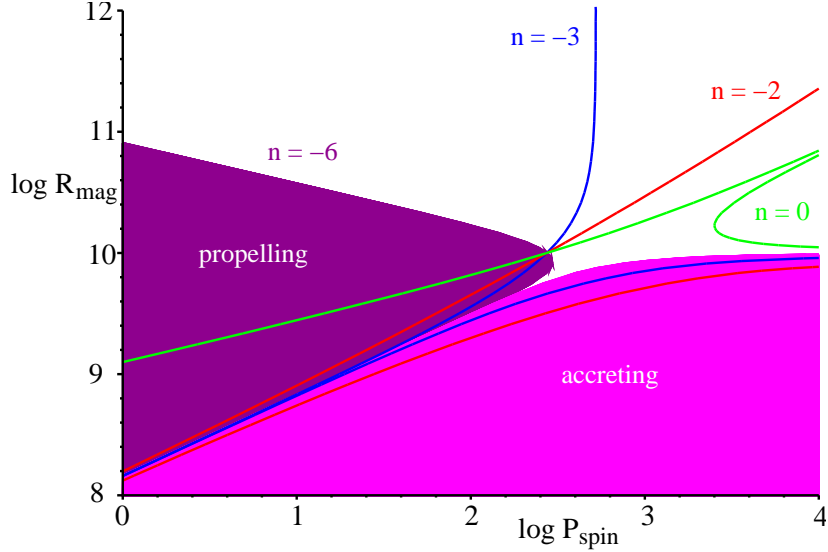}
    \caption{Spin-dependent magnetospheric radius for various radial
      powers $n$ in eq.~(1). Shaded areas indicate in the case $n=-6$
      a propelling and accreting branch, separated by the corotation 
      radius. As the effective $\rmag$ is the larger of the two at
      any given spin period, we find (for $n < -3$) a maximal spin
      period for which a propelling solution can exist.}
\end{figure}
\begin{figure}
    \plotone{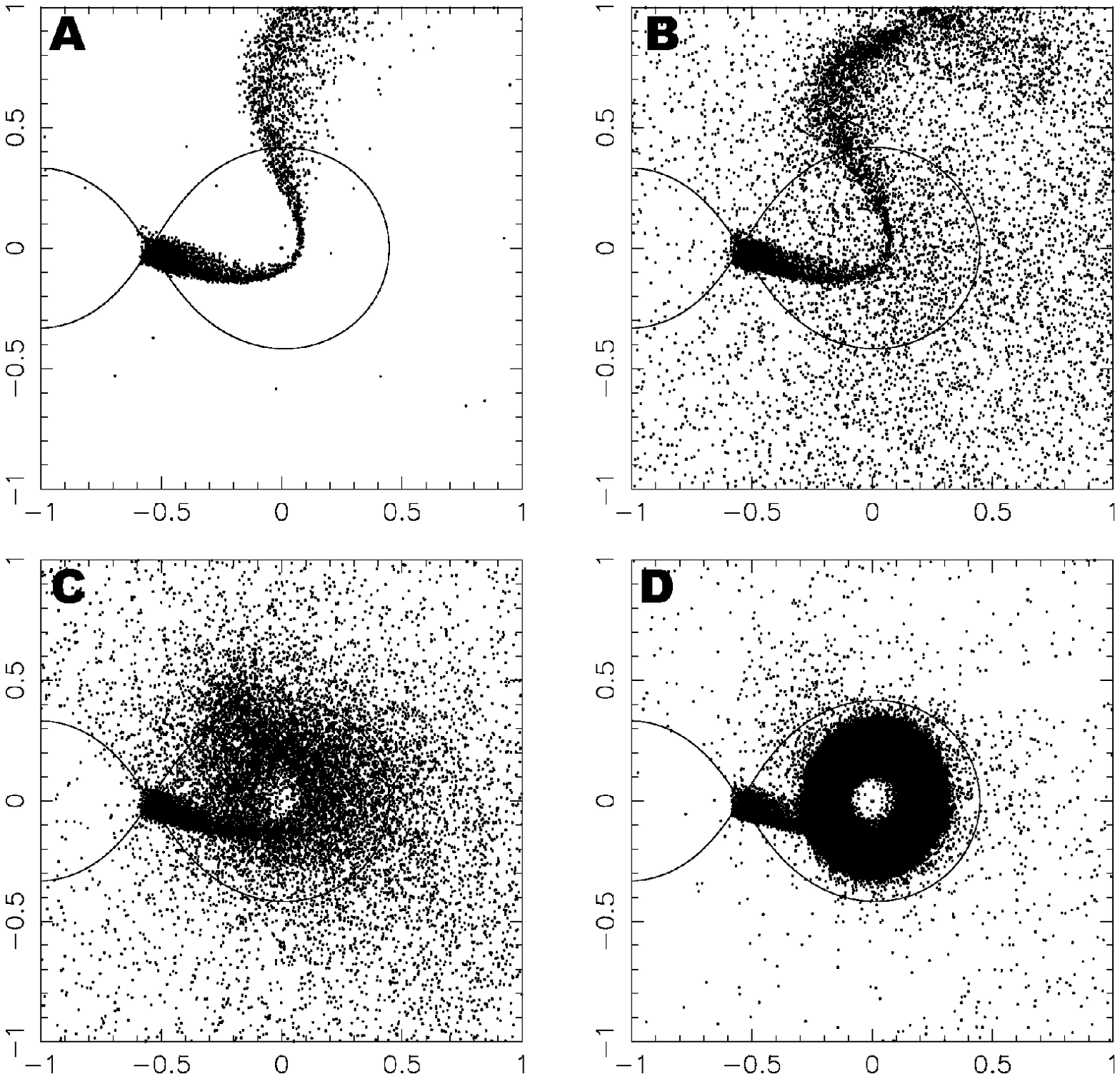}
    \caption{Snapshots of AE~Aqr at spin periods of $30$, $60$, $90$
      and $120 \, {\rm s}$ (panels A through D), corresponding
      almost linearly with time on a timescale of 
      $10^{7} \, {\rm yr}$. The current strong propeller will continue
      to work until mass ejection ceases and a disk forms with a large
      hole in the centre caused by a weak propeller. Accretion onto
      the WD will mainly occur during outbursts, when in a stable spin
      configuration the WD will regain all the angular momentum lost
      between outbursts.}
\end{figure}
Extending the usual definition of the magnetospheric
radius (e.g.\ Warner, 1995; Frank et al., 2002) we compare
the timescale of the magnetic interaction to the dynamical timescale
(in the disk-less case, I) and to the viscous timescale (in the
truncated disk case, II). Fig.~3 shows the resulting critical radii 
(shaded for $n=-6$, and an additional set of lines for $n=0,-2,-3$).
For $n=-6$ there is a critical spin period
\begin{equation}
    P_{\rm max} = 18 \, {\rm s} \,
        \mu_{33}^{2/3} \, \mdot_{16}^{-1/3} \,
    	m_1^{-2/3} \, \sigma_{9}^{1/3} \, c_{6}^{1/3}
\end{equation}
marking the maximum of the propelling branch in case I. 
Such a maximum exists for all $n<-3$ for both cases.
Additionally the asymptotic radius of the accreting branch in the
limit of slow rotation can be written as 
\begin{equation}
    R_{\rm asym} = 1.6 \times 10^{9} \, {\rm cm} \,
        \mu_{33}^{4/9} \, \mdot_{16}^{-2/9} \,
    	m_1^{-1/9} \, \sigma_{9}^{2/9} \, c_{6}^{2/9}
    \, ,
\end{equation}
which corresponds to the radius where the stream becomes threaded in
an ordinary polar. 
Both of these expressions describe case I where the mass
transfer stream is specified by 
$\mdot$ (in $10^{16} \, {\rm g \, s^{-1}}$), a stream width 
$\sigma$ (in $10^9 \, {\rm cm}$), and the sound speed at the L1 point
$c$ (in $10^6 \, {\rm cm \, s^{-1}}$). The WD 
mass is $m_1$ (in $\msun$) and its magnetic moment is $\mu$ (in 
$10^{33} \, {\rm G} \, {\rm cm^3}$).
The corresponding expressions for case II are 
\begin{equation}
    P_{\rm max} = 124 \, {\rm s} \,
    	   \alpha^{-18/125} \, \mu_{33}^{24/25} \, \mdot_{16}^{-33/125} \,
	   m_1^{-32/25}
\end{equation}
and
\begin{equation}
    R_{\rm asym} = 5.9 \times 10^{9} \, {\rm cm} \,
    	   \alpha^{-12/125} \, \mu_{33}^{16/25} \, \mdot_{16}^{-22/125} \,
	   m_1^{-13/25}
    \, ,
\end{equation}
with $\alpha$ describing the usual disk viscosity. 

Spin equilibria IPs are achieved when torques on the WD
cancel. Comparing $\rmag$ to various other important radii 
identifies such states (Warner, 1995; King \& Wynn, 1999). 
We are now able to analyse the various spin equilibria in 
mCVs provided we have good measurements of the periods
(easy), magnetic moments (difficult), masses and mass transfer rates
(even more so).

\section{The Future of AE~Aqr}

\subsection{Truncated disk due to weak propeller?}

Results of our most recent simulation are presented in Fig.~4.
Extending previous attempts (Wynn et al., 1997) to model the current
status of AE~Aqr, this calculation follows the spin
evolution through the next few $10^{7} \, {\rm yr}$.
During this time the spin-down continues at roughly the current rate.
We can clearly see that at $\pspin > 100 \, {\rm s}$ a 
truncated disk forms with a large inner hole of $R_{\rm in} \sim 2
\times 10^{10} \, {\rm cm}$. 

We compare this result in Fig.~5 to an analysis using critical radii.
In the left-hand panel the situation for AE~Aqr in its current 
state is shown: at the observed spin period of 33 sec and a strong
propeller (lower curve, as labelled), $\rmag$ is large enough to prevent disk
formation in the circularization region. As the WD slows down, a disk
will form and the effective $\rmag$ will switch to the truncated disk
case (pair of curves for hot and cold $\alpha$). As long as
this happens at $\pspin < P_{\rm max}$ (given by eq.~(4)), the next
stage of spin evolution is a weak propeller in a truncated disk.
In the final state of the computation in Fig.~4 we indeed find $R_{\rm
in} \approx \rmag$. 
For $P_{\rm max,hot} < \pspin < P_{\rm max,cold}$ a
stable equilibrium is possible where the
propeller slows the WD down during quiescence, 
while it is spun up by an equal amount during outbursts. 
This equilibrium can be maintained over secular timescales. 
Such a system will not be visible as a mCV during its rather long
quiescent stage between outbursts.
We can now construct a sequence of diagrams for AE~Aqr with
parameters taken from the post-TTMT evolution by Schenker et
al.\ (2002). The right-hand panel of Fig.~5 (showing only the
truncated disk pair of curves) is taken at $\porb = 5 \, {\rm h}$:
the $\rmag$ curves are shifted upwards along the corotation radius,
i.e.\ the central hole has grown to the point that spin equilibrium
is impossible. Further slow down of the WD ends the weak propeller
phase, and {\em for the first time} AE~Aqr will appear as a normal IP:
with a typical $\pspin / \porb$ ratio and $\porb$ in the range 
where the bulk of IPs are found (Fig.~2). 
\begin{figure}
    \plottwo{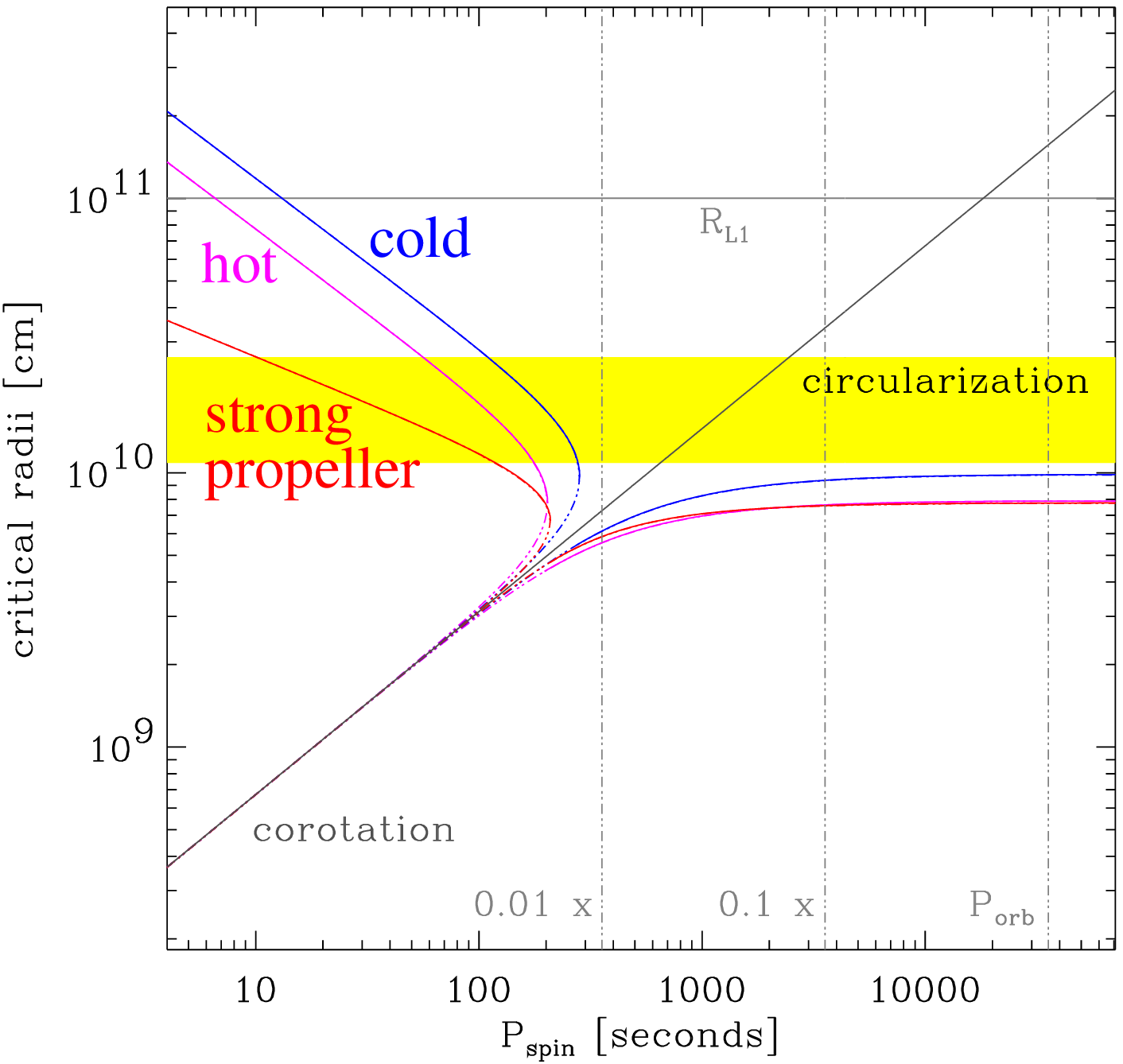}{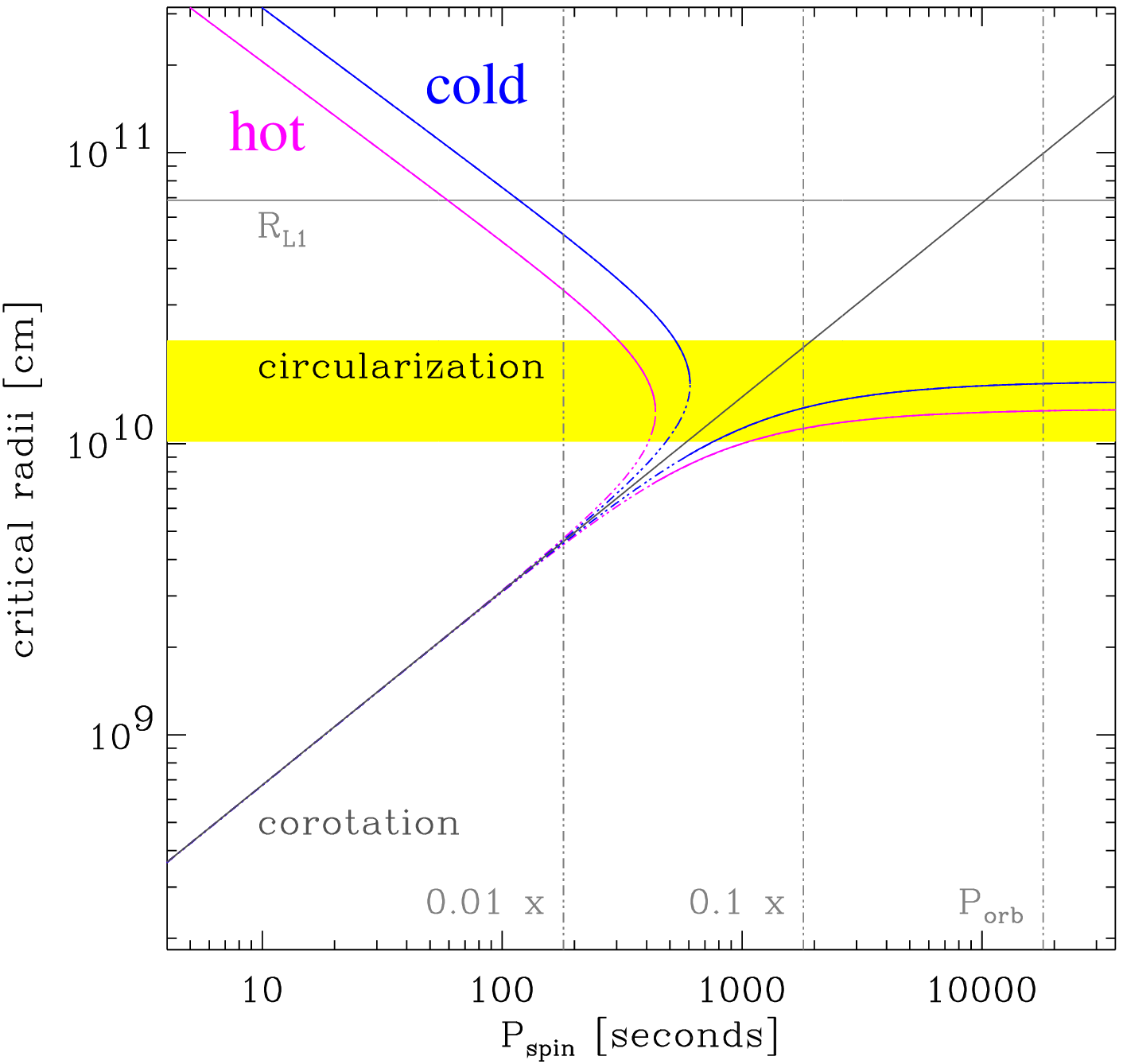}
    \caption{Critical radii of AE~Aqr at its current state with 
      $\porb = 9.88 \, {\rm h}$ (left panel) and 
      $\porb = 5 \, {\rm h}$ (right panel). See text for discussion.}
\end{figure}

\subsection{Stable strong propeller??}

We also identified a potentially stable configuration for a strong
(disk-less) propeller. The velocity of a stream reaches its maximum
value near the point of closest approach to the WD, the point at which
the rotational velocity of the magnetic field is lowest. So
for given $\porb$ and $q$ a critical spin period $P_{\rm crit}$ can be
found at which the sign of the drag force between the stream and
the rotating WD field changes. As the force depends strongly on radius
even a small inversion may be sufficient to balance the angular
momentum transferred from the WD to the stream while further away. This
may lead to a stable propeller, i.e.\ a situation where the mass
transfer stream is leaving the system yet {\em the WD spin does not
change}. In order to test this new idea, 
we have performed SPH simulations to investigate the possibility of a 
stable strong propeller.
As can be seen in Fig.~6, it is possible to construct
a strong propeller at this critical spin state.
Near the point of closest approach the velocity shear reverses and
the stream returns angular momentum to the WD. 
However, it would appear that this can only happen at spin periods
well above $P_{\rm max}$ found for AE~Aqr.

\section{Conclusion}

\begin{figure}
    \plottwo{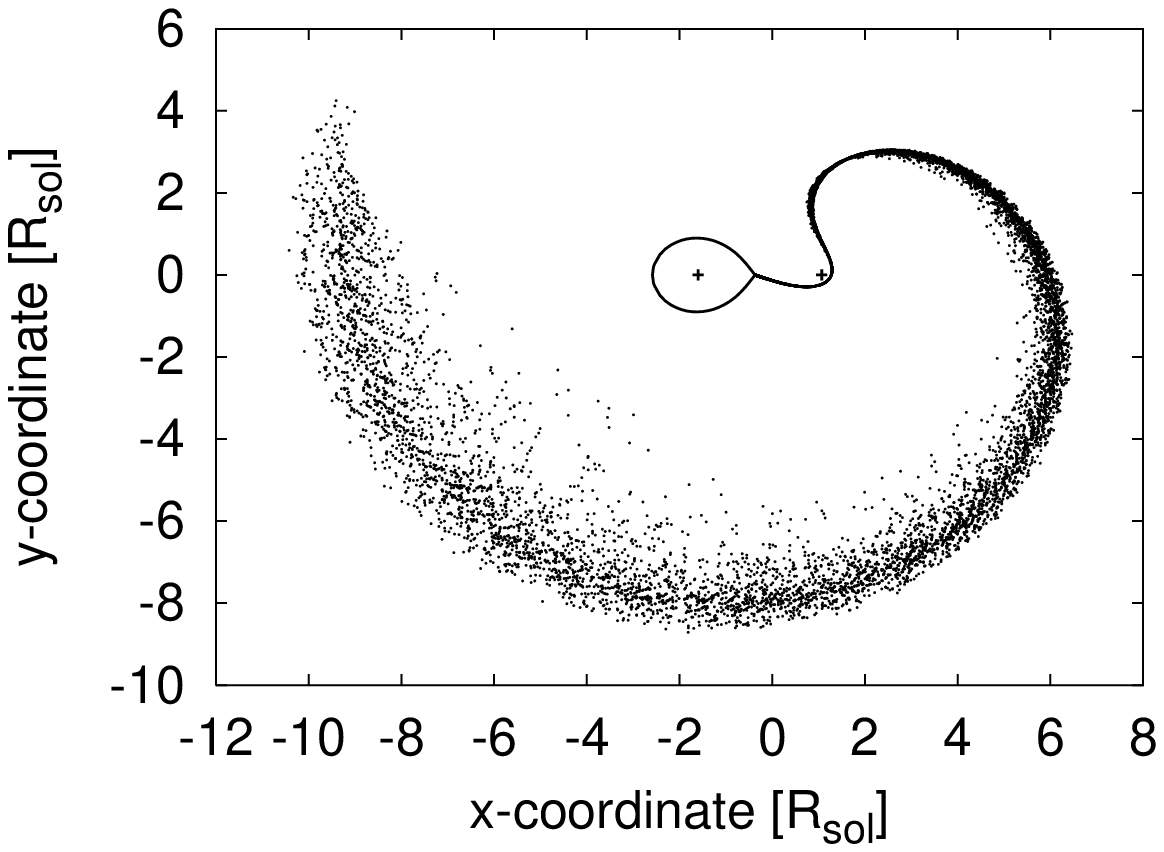}{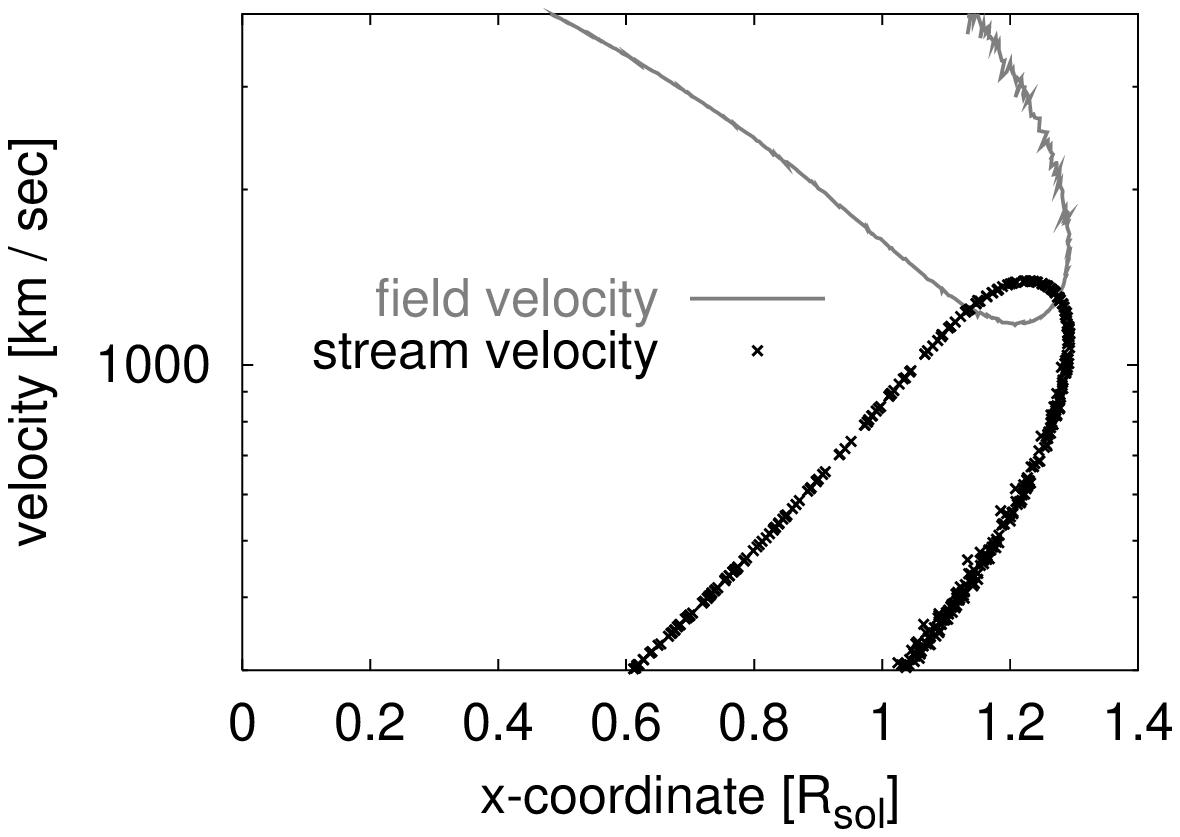}
    \caption{SPH simulation of a propeller with 
      $\pspin \simeq P_{\rm crit}$.
      In order to still have a strong propeller at this long spin
      period (left panel) a larger dipole moment is required than
      would be possible for the WD in AE~Aqr.
      Note how around the point of nearest approach the shear between
      the tangential velocity components changes sign (right panel).}
\end{figure}
In summary both the critical radius analysis and numerical
work indicate that AE~Aqr will continue to be a
strong propeller for a short time. Afterwards it is likely to
become a weak propeller system with a truncated disk.
At least theoretically, a stable strong propeller is possible,
although most probably not feasible for AE~Aqr. 
As long as the system can avoid accreting in a stable, IP-like
manner, any descendant of AE~Aqr may easily be overlooked as a 
mCV.
		      
Overall we conclude that TTMT evolution forces us to 
re-evaluate ideas about magnetic binary
evolution. The various spin states of magnetic WDs can 
lead to drastically different behaviour of otherwise similar systems. 
In particular, stable states may exist with or without a
disk, propelling or accreting. Not all of these states would currently
be considered to be magnetic systems on observational grounds.
      



\end{document}